\documentstyle[twoside]{article}

\begin{document}
${}$\\
\vspace{1.5 cm }
\begin{center}
{\large \bf The superconducting state in a single $\rm CuO_2$
layer: Experimental findings and scenario }
\end{center}
\vspace {0.10 in}
\begin{center}
{\normalsize  Rushan Han, Wei Guo }
\\
{\small \it School of Physics, Peking University, Beijing 100871 }
\end{center}
${}$\\
\begin{center}\begin{minipage}{4.5in}
{\small The notable results shown by Bozovic et al. from
measurements of synthesized one-unit-cell thick of $\rm
La_{1.85}Sr_{0.15} Cu O_4$, $\rm La_2 Cu O_4$ HTS/AF/HTS tri-layer
junctions query the predictions by prevalent models since there is
no observable mixing of superconductivity and anti-ferromagnetism
between the layers. In this article we make a brief survey of
experimental results on the electronic structure of high-$T_c$
cuprates and magnetic properties. Based on our analysis, we
attribute superconductivity of a $single$ $\rm CuO_2$ layer to
spin pairing via local exchange interactions. }
\end{minipage}
\end{center}
${}$\\
${}$
${}$\\
${}$\\
Recently, Bozovic et al. synthesized one-unit-cell thick
superconducting LSCO, insulating LCO HTS/AF/HTS tri-layer
junctions in a way of stacking digitally by using advanced
Molecular Beam Epitaxy technology [ Nature 422, 873(2003) ]. The
result shown by Bozovic et al. indicating no mixing of
superconductivity and anti-ferromagnetism in HTS/AF/HTS layers
queries the predictions by prevalent models. The fact that
superconductivity holds in one-unit cell thick LSCO layer also
questions the idea of the inter-layer coupling as a pairing
mechanism. Bozovic et al. have shown that there is 1 eV energy
difference between LSCO and LCO, and the quasi-two-dimensional
mid-gap states in LSCO, i.e. the energy position of the
superconducting state is in between the empty and the valence band
of LCO. This observation agrees with some earlier works by other
investigators: the normal states of charge carriers are in between
two energy bands of LCO, the chemical potential pins in the middle
of the gap rather than at the bottom. The doped holes appeared in
oxygen band activate $p$ electrons, the density of itinerant
electrons is found to be proportional to concentration of the
doped holes $x$ ( also the density of superfluids $n_s$ ). Mixing
between orbital $2p$ (O) and $3d$ (Cu) has been observed, the
mixing ratio is 8 : 2. The mixed states are not located in Cu site
or O site, but extended itinerant states with metallic mobility.
The superfluid carriers are mainly $p$ electrons rather than $d$
electrons.

High-$T_c$ phenomena occur in a special magnetic structure only.
Among the huge number of compounds belong to the family of
strongly correlated systems, non-Fermi-liquids, d-f metals and
perovskite type materials, high-$T_c$ cuprate is the only one
superconductive with high-transition temperature. On the other
hand, all high-$T_c$ cuprates with different compositions exhibit
universal physical properties such as the phase diagram,
non-Fermi-liquid behavior, etc. The superconducting phase emerges
in a doping region between insulting ( charge transfer type ) and
Fermi liquid regimes, the transition temperatures vary with the
doped hole concentration $x$ companied with variation of the
number of charge carriers and short range order of
anti-ferromagnetism. Itinerant $p$ electron also plays a role of
mediation for the superexchange between Cu ions ( in Wannier
representation ), the physical properties of charge carriers ( $p$
electrons ) and variations of magnetic correlation between local
Cu spins influenced by hole doping should be carefully examined
since the nearly free local spins created by holes in the
anti-ferromagnetic background may interact with charge carriers
via Kondo exchange. In the following we list main experimental
results on the electronic structure of high-$T_c$ cuprates with
brief comments. The article is divided into four parts: i. The
physical properties of charge carriers, ii. Varying of
anti-ferromagnetic background by changing the doped hole
concentration, iii. Interactions between charge carriers and the
anti-ferromagnetic background, iv.
Physical properties of the superconducting state.\\

\noindent
1. Bozovic I. et al. Nature 422, 873(2003). Evidence for
quasi-two-dimensional electronic state located in the middle of
band gap.\\
2. Uchida S. et al. PRB. 43, 7942(1991). Electronic state
in the middle of band gap shown by optical experiment.\\
3. Fujimori Lanl. 0011293. Evidence for electronic state in the
middle of band gap shown by transfer of spectral weight in ARPES.\\
4. Ino Lanl. 0005370. Evidence for electronic state in the middle
of band gap shown by transfer of spectral weight in ARPES.\\
5. Ino PRL. 79, 2101(1997). Evolution of chemical
potential, pining in the middle of band gap.\\
6. Romberg H. et al. Phys. Rev. B 42, 8768(1990). O1s
$\rightarrow$ O2p resonance excitation, evidence for holes in
oxygen sites.\\
7. Nucker N. et al. PRB. 37, 5158(1988); PRB. 59, 6619(1989).
Holes in oxygen sites.\\
8. Arko A. L. et al. et al. PRB. 40, 2268(1989). The mixing ratio
O2p : Cu3d $\sim $ 8 : 2\\
9. Feng D. L. et al. Science 289, 277(2000). Superfluid density
$n_s \propto T_c \propto x $\\
10. Yoshida T. et al. arXiv: cond-mat/0206469. Quasi-particle
peak, the number of charge carriers $n \propto x $ and metallic
behavior.\\

By increasing the doped hole concentration, the long
anti-ferromagnetic order of Cu moment in the $\rm CuO_2$ plane
turns into short range order, ultimately to disorder companied
with delocalization of $d$ electrons on the Cu sites. In a proper
doping range, $d$ electron with spin S =1/2 is localized, oriented
along b axis. The magnetic correlation between local spins in the
$\rm CuO_2$ plane is anisotropic Ising like. Doped holes in the
anti-ferromagnetic bonds suppress local spin correlation and yield
the incommensurate splitting of neutron scattering magnetic peak
and nearly free local spins. In case of dynamical phase
separation, anti-ferromagnetic background is imperfect, flipping
local spins create the "defects" in the anti-ferromagnetic
background ( these "defects" are not static since the holes hop
site to site ). This is another important effect brought by hole
doping, which is overlooked by many people. This factor plays an
important role in the variation of the physical properties of
high-$T_c$ cuprates with increasing doped hole concentration. In
heavy doping region $x > 0.19 $, when delocalization of $d$
electrons on Cu site begins, the density of charge carriers
changes from $x$ to $1-x$. On the lower side of this critical
doping value, there always exist local $\rm Cu^{2+}$ ions. Above
the critical doping level, superconductivity
vanishes with collapsed anti-ferromagnetic background.\\

\noindent 11.
Aeppli G. et al. "Lecture notes for E. Fermi Summer
School, Varenna", 1992. Short range anti-ferromagnetic correlation
length $\xi_s \sim 0.38/\sqrt x $ nm .\\
12. Birgeneau R. J. et al. "Physical Properties of High
Temperature Superconductivity I" , edited by Ginsberg D. M. (
World Scientific, Singapore, 1989 ), p154, Fig 1. Local spins
parallel to a main axis in the $\rm CuO_2 $ plane.\\
13. Lavrov A. N. et al. PRL. 87, 17001(2001). Anisotropy of
susceptibility, spins parallel to b axis. Evidence for the
existence of nearly free moments.\\
14. Dai Peng-Cheng et al. PRB. 63, 54525(2001). Incommensurate
splitting of magnetic $(\, \pi\:, \pi \,)$ peak persisting into
the superconducting state.\\
15. Yamada K. et al. PRB. 57, 6165(1998). Incommensurate splitting
$\delta \propto x $ saturating near the optimal doping.\\
16. Zhou X. J. et al. Science 286, 268(1999). Phase separation
shown by ARPES, proposed charge and magnetic ordering structure.\\
17. Lake B. et al. Nature 400, 43(1999). Incommensurate splitting,
energy of spin gap is independent of momentum.\\
18. Uchida S. et al. Physica C 282-287, 12(1997). Delocalization
of Cu ions. The density of charge carriers changes from $x$ to
$1-x$.\\
19. Brooks N. B. PRL. 87, 237003(2001). Detecting $\rm Cu^{2+}$
ions by using spin resolute photo emission technique. \\
20. Tallon J.L. et al. Physica C 349, 53(2001). So called "0.19"
problem, delocalization of local charge companied with variation
of physical properties.\\

It is observed that the scattering rate of electrons varies about
three orders of magnitude near transition temperature $T_c$,
electron-phonon interaction is not a main factor to the
superconducting transition. On the other hand, anomalous Hall
effect observed indicates electron-local-moment scattering. AHE is
characterized by linear temperature dependence of Hall number,
which is a common feature for many of magnetic materials. Smit and
Luttinger attribute AHE to skew scattering. AHE in Kondo-type
systems was first discussed by Fert et al.(not like dilute
magnetic impurity systems, local moments in high-$T_c$ cuprates
are correlated, no saturation at high temperatures). The
simultaneous suppression of superconductivity and the
superexchange by substitution of Cu by Zn and Ni indicates that
S=1/2 is required for high-$T_c$ physics. In the doping region ( $
0.02 < x < 0.15 $ ), the range of anti-ferromagnetic order is
short, the superconducting coherent length is comparable to the
magnetic correlation length. The mobility of charge carriers is
inversely proportional to anti-ferromagnetic correlation length
indicating the magnetic origin of microscopic interactions. Charge
carriers in the mid-gap state interacting with mobile "defects" in
the anti-ferromagnetic background is the key to superconductivity.\\

\noindent
21. Bonn D. A. et al. PRB. 47, 11314(1993). Scattering
rate $1/\tau$ varies three order of magnitude near transition temperature.\\
22. Zhang Y. et al. PRL. 84, 2219(2000). Measurement of
Wiedemann-Franz ratio indicate that electron-electron scattering
is dominating.\\
23. Ong N. P. "Physical Properties of High Temperature
Superconductivity II" edited by Ginsberg D. M. (World Scientific,
Singapore 1990 ), p460. Hall number varies linearly with temperature.\\
24. Bergmann G. Physics Today, 1979, p25. A general survey of the
origin of temperature dependent anomalous Hall effect, asymmetric
scattering between conduction electrons and local moments.\\
25. Luttinger J. M. Phys. Rev. 112, 739(1958). The concept of skew scattering\\
26. Fert A. et al. PRL. 28, 303(1972); Levy P M. et al. J. Appl.
Phys. 63, 3869(1988). Skew scattering in Kondo-type systems.\\
27. White P. J. et al. arXiv: cond-mat/9901349, 9901354. Zn substitution.\\
28. Alloud H. et al. PRL. 67, 3140(1991). Substitution of Cu by Zn
suppresses superconductivity and anti-ferromagnetism.\\
29. Kakurai K. et al. PRB. 48, 3485(1993). Substitution of Cu by
Zn suppresses superconductivity and anti-ferromagnetism.\\
30. Ando Y. et al. PRL. 87, 17001(2001). The mobility of the doped
holes is proportional to AF correlation length.\\

The $\rm CuO_2$ plane is a common structure unit of all high-$T_c$
cuprates. Bozovic et al show quasi-two-dimensionality of the
superconducting state and short proximity effect. In the phase
diagram, $T_c/T_c^{\rm max }$ is a parabolic function of $x$ near
the optimal doping, the curvature $\kappa $ and the optimal doping
$x_0$ are universal constants. Excitations in a high-$T_c$ cuprate
are complex, especially the pseudogap with $d_{x^2 - y^2 }$
symmetry and isotropic spin gap. The revival of spin gap
responding to external magnetic field reveals that the microscopic
in-plane interaction is magnetic, which is agreed with observed
time-reversal symmetry breaking effect in polarized photo-emission
experiment( need to be confirmed ). The intrinsic inhomogenity of
the superconducting state in atomic scale by using STM method
queries all proposed k space pairing models.\\

\noindent 31. Batlogg B. et al. Physica C 235-240, 130(1994). A
survey of the pseudogap.\\
32. Xu Zhuan et al. Nature 406, 486(2000). Vortex-like excitation
in underdoped samples above $T_c$ (100 $\sim $220 K ), evidence
for incoherent pairing.\\
33. Wen H. H. et al. arXiv: cond-mat/0301367\\
34. Tallon J. L.  et al. PRL. 75, 4114(1995); PRB. 51,
12911(1995). Universal expression for empirical $T_c$ law.\\
35. Tsuei C. C. et al. PRL. 73, 593(1994). Tri-crystal junction,
$d_{x^2 - y^2 }$ symmetry.\\
36. Rossat-Mignod J. et al. Physica C 185-189, 86(1991); Physica B
169, 58(1991). Spin gap $\Delta \sim T_c $.\\
37. Yamada K. PRL. 75, 1626(1995). Spin gap in LSCO.\\
38. Pan S. H. et al. Nature 413, 282(2001). Intrinsic microscopic
inhomogenity of the superconducting state.\\
39. Ginsberg D. M. "Physical Properties of High Temperature
Superconductivity I", edited by Ginsberg D. M. (World Scientific,
Singapore 1989 )p1. Short superconducting coherent length.\\
40. Kaminski A. et al. Nature 416, 610(2002). Polarized
photo-emission, time-reversal symmetry breaking in the pseudogap
and the superconducting state.\\
41. Lake B. et al. Science 291, 1759(2001). Indication of the
magnetic origin of microscopic interaction.\\

Two interactions are essential to the superconducting state in a
single $\rm CuO_2$ layer: the superexchange ( coupling constant K
$\sim $ 0.1 eV ) between $\rm Cu^{2+}$ions and the Kondo exchange
between itinerant charge carriers and "defects" in the
anti-ferromagnetic background i.e. nearly free spins ( coupling
constant J $\sim$ 0.01 eV ). We interpret the coexistence of
anti-ferromagnetism and superconductivity as following: $p$
electron being the mediation for the superexchange between Cu
spins is activated by hole doping breaking the long range order of
Cu spins. On the other hand, electron spin coupling mediated by
nearly free Cu spins via the Kondo exchange gives rise to electron
pairing in spin sector. A notable result of spin pairing is that
the wavefunction of two opposite spins is symmetric rather than
anti-symmetric( susceptibility measurement in NMR and neutron
scattering experiment can not determine the symmetry of spin
wavefunction ). The projection of the orbital wavefunction of the
electron pair in the $\rm CuO_2$ plane has the $d_{x^2 -y^2}$
symmetry which is forced by spin and crystal symmetries. Orbital
wavefunction along $c$ axis is restrained as shown by Bozovic et
al. As explained by the K-J model, the two dimensionless
quantities $\kappa = 86.2 $ and optimal doping $x_0 = 0.16 $ in
the universal parabolic function of $T_c/T_c^{\rm max } $ are not
independent, one relates to another determined by two fundamental
constants $K_0$( for undoped sample )and $J$. The K-J model also
explains: the upper limit of $T_c^{\rm max } \sim 150K $,
delocalization of $d$ electron at $x \approx 0.2$, short range
magnetic order, spin gap and the pseudogap, low superfluid density
$n_s \propto  x \propto T_c $, time-reversal symmetry breaking and
non-fermi-liquid behavior, etc.\\

\noindent
42. Guo W. et al. arXiv: cond-mat/0303155; Physica C 364-365,79(2001).
The K-J model.\\

By summary, we have listed essential literatures of experimental
results showing the electronic structure of high-$T_c$ cuprates (
limited by article length we have omitted many important works ).
To understand high-$T_c$ superconductivity, we have to put every
piece of experimental findings together to figure out the hidden
scheme like a puzzle game. Based on our analysis, we attribute
superconductivity of a single $\rm CuO_2 $ layer to spin pairing
via local exchange interactions.

\end{document}